\documentclass[prb,twocolumn,showpacs,preprintnumbers,amsmath,amssymb,showkeys]{revtex4}
\usepackage{amscd,graphicx,psfrag}

\newcommand{\hvect}[1]{\hat{#1}}
\newcommand{\vect}[1]{\vec{#1}}
\newcommand{\vectsym}[1]{\vec{#1}}

\newcommand{\avg}[1]{\left\langle #1 \right\rangle}

\newcommand{\alphaDTconv}{\alpha_\text{DT}^\text{cont}}
\newcommand{\betaDTconv}{\beta_\text{DT}^\text{cont}}
\newcommand{\zDTconv}{z_\text{DT}^\text{cont}}
\newcommand{\alphaWVconv}{\alpha_\text{WV}^\text{cont}}
\newcommand{\betaWVconv}{\beta_\text{WV}^\text{cont}}
\newcommand{\zWVconv}{z_\text{WV}^\text{cont}}
\newcommand{\alphaWV}{\alpha_\text{WV}^\text{\phantom{cont}}}
\newcommand{\betaWV}{\beta_\text{WV}^\text{\phantom{cont}}}
\newcommand{\zWV}{z_\text{WV}^\text{\phantom{cont}}}

\begin{document}

%\preprint{cond-mat/XXX}

\title{%
Growth instability due to lattice-induced topological currents in limited mobility epitaxial growth models}

\author{Wittawat Kanjanaput}
\email{w.kanjanaput@gmail.com}
\author{Surachate Limkumnerd}
\email{surachate.l@chula.ac.th}
\author{Patcha Chatraphorn}
\email{patcha.c@chula.ac.th}
\affiliation{%
Department of Physics, Faculty of Science, Chulalongkorn University,\\
Phayathai Rd., Patumwan, Bangkok, 10330, Thailand\\
Research Center in Thin Film Physics, Thailand Center of Excellence in Physics,\\
CHE, 328 Si Ayutthaya Rd., Bangkok 10400, Thailand
}

\date{\today}

\begin{abstract}
The energetically driven Ehrlich--Schwoebel (ES) barrier had been generally accepted as the primary cause of the growth instability in the form of quasi-regular mound-like structures observed on the surface of thin film grown via molecular beam epitaxy (MBE) technique. Recently the second mechanism of mound formation was proposed in terms of a topologically induced flux of particles originating from the line tension of the step edges which form the contour lines around a mound. Through large-scale simulations of MBE growth on a variety of crystalline lattice planes using limited mobility, solid-on-solid models introduced by Wolf--Villain and Das Sarma--Tamborenea in 2+1 dimensions, we propose yet another type of topological uphill particle current which is unique to some lattice, and has hitherto been overlooked in the literature. Without ES barrier, our simulations produce spectacular mounds very similar, in some cases, to what have been observed in many recent MBE experiments. On a lattice where these currents cease to exist, the surface appears to be scale-invariant, statistically rough as predicted by the conventional continuum growth equation.
\end{abstract}



\maketitle

\section{Introduction}
The growth of crystalline thin films via molecular beam epitaxy (MBE) has attracted several interests experimentally\cite{Cham00,Arth02} due to ever growing applications in many fields, and theoretically\cite{BaraStan95,PimpVill98} due to its rich surface structures. Incoming flux of atoms deposit onto a substrate in the layer-by-layer fashion generally makes the film completely free of defects. Excess energy of the adatoms allows them to diffuse along the surface away from their initial landing positions. These atoms tend to minimize their energy by moving towards sites with high coordination number such as those along island step edges. The process produces an instability in the growth morphology leading to the formation of surface structures.

Over the past decades, there have been numerous experimental evidences of a pyramid-like mound morphology with a well-defined mound shape and a selective slope.\cite{ErnsFabrFolkLapu94,ColuCottMendLandCarv98,VanNostCheyCahi98,KalfSmilComsMich99,FinnHomm99,FausReusShumWeinTheiGold00,NeelMaroDouiErns03} The origin of such a structure has mainly been attributed to the presence of Ehrlich--Schwoebel (ES) barrier~\cite{EhrlHudd66,SchwShip66,SchwShip69,Vill91} which, according to Burton--Cabrera--Frank theory,\cite{BurtCabrFran51,Krug97} hinders atoms from moving down a terrace resulting in the tendency for them to prefer to flow ``uphill.''
Several investigators have confirmed mound formation due to ES barrier through their computer simulations.\cite{PoliVill96,CamaCrosCapiAlvaFerr99,SchiKohlReen00,LiuShen05} 
%More recently another type of destabilizing ES step-edge current was discovered which causes atoms to flow along a step edge within the same terrace toward a kink site or over a corner site.\cite{MurtCoop99,PierDorsEins99}
More recently another type of destabilizing ES step-edge current was discovered which occurs from the imbalance between the flow of atoms along a step edge within the same terrace toward a kink site and the flow toward the kink site over a corner.\cite{MurtCoop99,PierDorsEins99}
The kink ES current, unlike the former kind, causes the motion within the same terrace, and thus can only occur on a surface with spatial dimension higher than one.  Until the past decade, it was believed that ES barrier was the sole cause of mound morphology. One of the authors \cite{ChatToroSarm01} was able to obtain a mound-like structure without implementing the ES barrier. Unlike the ES mechanism which are energy-assisted, this mounding instability is probabilistic and topological in nature. A net current arises via the unbalanced between atoms diffusing up and down a step edge, hence the name ``step-edge diffusion'' (SED) current. On a simple cubic lattice, this type of current only occurs around a corner or a kink site which is similar to the kink ES current. The mounding instability through SED current did not emerge spontaneously, and was observed only after the use of the so-called ``noise reduction technique'' to suppress the deposition and nucleation noise.\cite{SarmPunyToro00,ChatToroSarm01,RangChat06}
It is unclear that SED mechanism, initially studied in a simple cubic system, always occurs, and always leads to mounding structure in \emph{all} crystalline lattice structures. Do other topological currents exist in other structurally different crystalline lattices?

Aside from a qualitative description of mound shapes, surface morphology is quantitatively measured in terms of its roughness as defined by
\begin{equation}\label{E:roughness}
	W(L,t) \equiv \left[\avg{h^2(t)}-\avg{h(t)}^2 \right]^{1/2},
\end{equation}
where $\avg{h^n(t)} \equiv L^{-d}\sum_i h^n_i(t)$ is the average of atomic heights (to the $n^\text{th}$ power) over all lattice sites of dimension $d$ and lateral size $L$. It is believed that $W(L,t)$ exhibits a power-law scaling behavior of the form,
\begin{equation}
	W(L,t) \sim L^\alpha f(\xi(t)/L)\,,
\end{equation}
where the scaling function $f(x)$ and the lateral dynamical correlation length $\xi(t)$ are given by
\begin{equation}\label{E:scalingFn}
	f(x) \sim \begin{cases} x^\beta\,,\quad x\ll 1, \\
								1\phantom{^\alpha}\,,\quad x \gg 1,
				\end{cases}\, \text{and} \quad \xi(t)\sim t^{1/z}
\end{equation}
with $\alpha$, $\beta$, and $z=\alpha/\beta$, as suggested by the dynamic scaling theory, being the roughness, growth, and dynamical exponents respectively. In 1+1 dimensions (one spatial dimension + one time), These exponents directly associate a given discrete growth model to a universality class. Despite impressive success in 1+1 dimensions, the computational results of several toy models do not conform to the theoretical predictions in 2+1 dimensions. The presence of mound morphology, a missing feature in one spatial dimension, suggests that a given model may belong to a different universality class in a different dimension, thus rendering the universality class concept futile.\cite{ChatSarm02,SarmChatToro02,OlivReis07} 

% Thesis statement
In this paper we propose, in addition to ES barrier, a competing mechanism for mound formation as a consequence of probabilistic terrace currents due to the geometry of a film's crystalline structure.
We begin, in Section~\ref{S:models}, by describing the helical boundary conditions essential for constructing representations of various crystal structures. Growth simulations are then performed on each of these structures according to a set of diffusion rules. Calculations of surface roughness and the critical exponents are carried out in Section~\ref{S:calculations}. The roughness exponent, in particular, implies the existence of mound morphology on a particular crystalline thin film. In Section~\ref{S:currents}, we work out the probabilistic currents of a few crystal structures for illustrative purposes. In addition to the SED current, a new type of topological current is discovered which also causes particles to flow in the uphill direction. We discuss the mound formation mechanism in connection with the underlying continuum growth equation in Section~\ref{S:discussion}, and summarize our work in Section~\ref{S:conclusion}.

\section{Models and crystal structures}\label{S:models}
\begin{figure}[htb]
	\centering
    \includegraphics[width=.48\textwidth]{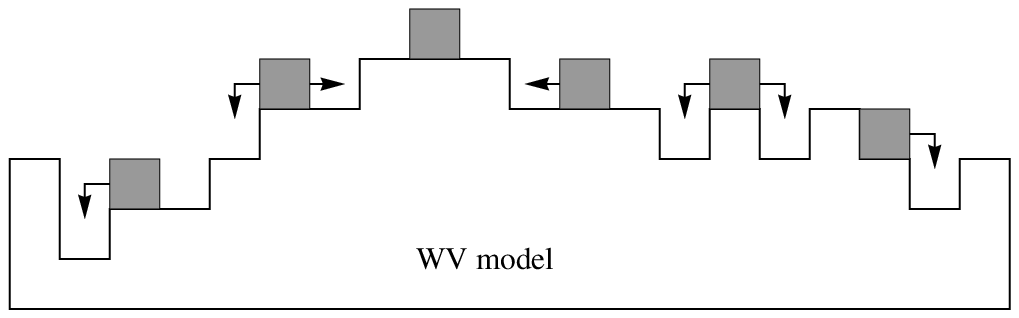}\\
    (a)\\ \vspace{.2in}
    \includegraphics[width=.48\textwidth]{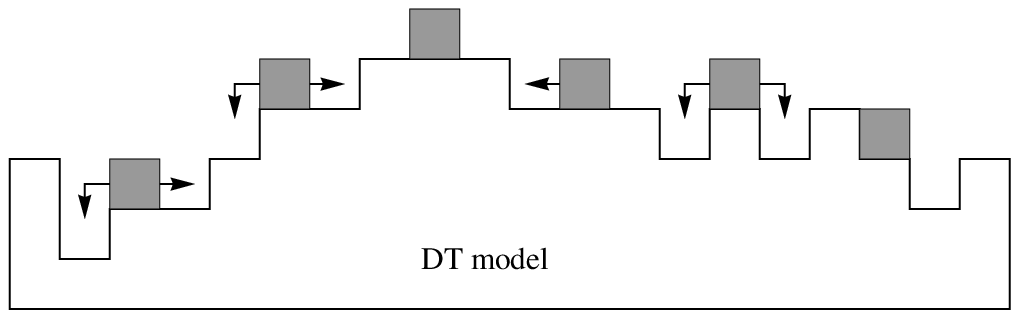}\\
    (b)
    \caption{One dimensional (a) Wolf--Villain model where an atom moves to maximize its coordination number and (b) Das Sarma--Tamborenea model where an atom moves to increase its coordination number given that the current bond count is less than two. An atom falls on one of the shaded regions can move in one of the directions specified by an arrow.}
    \label{fig:DiffRules}
\end{figure}

% XXX Too strong?
Limited mobility solid-on-solid diffusion models as means to model the growth of epitaxial thin film continue to attract many interests despite its simplicity and decades of intense investigations both computationally and analytically.
In these models, boxes representing atoms are sprinkled down from atop and subsequently relax to their final atomic positions without any voids or overhangs according to a given set of rules. We examine the limit of low substrate temperature so that deposition atoms can move at most to one of the nearest neighboring sites before coming to rest. Three most prominent diffusion rules are (i) Wolf--Villain~\cite{WolfVill90} (WV) where adatom moves to \emph{maximize} its nearest neighbor bonds, (ii) Das Sarma--Tamborenea~\cite{SarmTamb91} (DT) where adatom moves to \emph{increase} its nearest neighbor bonds provided that the current number of bond is less than two, and (iii) Edward--Wilkinson~\cite{EdwaWilk82} (EW) where adatom moves to the nearest neighbor with the minimal local height. Conventionally most modelling works are performed on two-dimensional rectangular lattices making them only applicable to simple cubic materials. At this stage, we have chosen to experiment with only WV and DT models and disregard EW model because we do not believe that height minimization would lead to any structural formation. %% XXX Check this!
Figure~\ref{fig:DiffRules} gives a schematic diagram of WV and DT diffusion rules in one spatial dimension.

%%% Figure showing various diffusion rules.

To overcome complications in representing any crystalline structures, we adopt the helical boundary conditions~\cite{NewmBark01} which represents any $d$-dimensional lattice using a one-dimensional chain. As an example, a site on an $M \times N$ rectangular lattice with coordinates $(i,j)$ is located at the $(iN+j)^\text{th}$ element of the chain of length $MN$ (counting from zero). Its four nearest neighbors at $(i\pm 1,j)$ and $(i,j\pm 1)$ are mapped to element number $(i\pm 1)\! \mod MN$ and $(i\pm N)\! \mod MN$ respectively. Other structures can be constructed in a similar manner. The main advantage of the use of helical boundary conditions over the conventional ones lies in the flexibility in representing any $n$-dimensional structure using one-dimensional chain of an arbitrary length (not restricted, e.g., to integer$\times$integer in the case of simple cubic crystal) resulting in the simplicity of the simulation code.

In this work we simulate the growth of 6 crystal structures on 7 planes: simple cubic (SC), body-centered cubic (BCC), face-centered cubic (FCC) on (001) and (111) planes, simple hexagonal (SH), ideal hexagonal closed pack (iHCP) where the ratio $c/a = \sqrt{8/3}$, and hexagonal closed pack (HCP) where $c/a < \sqrt{8/3}$. The structures whose substrate plane is not listed are understood to be on the (001) plane, or (0001) in the case of the three hexagonal lattices.
In all simulations we assume that atoms in the substrate and the film are of the same type so that the film's crystal orientation is the same as that of the substrate and no stress of any kind is produced along the interface. Unlike other conventional simulations where time is measured in units of monolayers, here it is measured in unit-cell layers (UL) since each structure viewed from a different plane may contain a different number of atomic layers per unit cell.

%%%%%%%%%%%%%%%%%%%%%%
% Exponents & Mounds %
%%%%%%%%%%%%%%%%%%%%%%
\section{Critical exponents and the existence of mound morphology}\label{S:calculations}
Unlike the generic non-equilibrium surface growth where voids and defects are prevalent, the study of kinetic surface roughening of non-equilibrium growth models of the solid-on-solid type remains a subject under much scrutiny.\cite{BaraStan95,PimpVill98} The former has been generally accepted as belonging to the Kardar--Parisi--Zhang universality class,\cite{KardPariZhan86} originated in the study of Eden cluster growth\cite{Eden61} and diffusion-limited aggregation,\cite{WittSand81,BazaChoiDavi03} with the dominant growth direction pointing along the surface normal, giving rise to a term proportional to $\sim (\vectsym{\nabla} h)^2$. Solid-on-solid growth models, replicating MBE growth, offer much richer behaviors especially when the spatial dimension is higher than one. It has been known that a slight change in a diffusion rule results in an alteration of the universality class. In 1+1 dimensions, WV and EW both belong to EW universality class (with $\alpha = 1/2$, $\beta = 1/4$, $z=2$) while DT model belongs to MBE class (with $\alpha = 1$, $\beta = 1/3$, $z=3$) despite closer diffusion rule to WV than is EW model.\cite{BaraStan95} Based on these results, together with some symmetry arguments,\cite{BaraStan95} it can be shown that both diffusion rules obey the following continuum growth equation:
\begin{equation}\label{E:ContEqn}
	\frac{\partial h}{\partial t} = \nu_2 \nabla^2 h - \nu_4 \nabla^4 h + \lambda_{13}\vectsym{\nabla}\cdot(\vectsym{\nabla} h)^3 + \lambda_{22}\nabla^2(\vectsym{\nabla} h)^2 +\eta
\end{equation}
with $\nu_2 = 0$ for DT model by symmetry, and $\nu_2$ is very small compared with $\nu_4$ and $\lambda_{22}$ but positive for WV model. The term proportional to $\vectsym{\nabla}\cdot(\vectsym{\nabla} h)^3$ is often neglected (by setting $\lambda_{13}$ to zero) because it generates the $\nu_2 \nabla^2 h$ term upon renormalization. Physically speaking this term (with $\lambda_{13}>0$) gives a dissipative effect similar to the $\nu_2\nabla^2 h$ term, but at a shorter length scale. The stochastic nature is captured by the Gaussian noise $\eta(\vect{x},t)$ where $\avg{\eta(\vect{x},t)} = 0$, and
\begin{equation}
	\avg{\eta(\vect{x},t)\eta(\vect{x}',t')} = 2D\,\delta(\vect{x}-\vect{x}')\delta(t-t').
\end{equation}

Provided the validity of Eq.~(\ref{E:ContEqn}) with the coefficients having the same sign as in 1+1 dimensions, the DT exponents in 2+1 dimensions are found to be $\alphaDTconv = 2/3$, $\betaDTconv = 1/5$, and $\zDTconv = 10/3$.\cite{BaraStan95} The WV model in this dimension is predicted to produce logarithmically smooth surface with $\alphaWVconv = \betaWVconv = 0$ ($\text{log}$)), while the dynamical exponent still obeys power law scaling with $\zWVconv = 2$. A large-scale simulation on a SC substrate by Das Sarma~\emph{et al.}, however, gives a contradictory result.\cite{SarmChatToro02} They reported that the DT model behaved as if it were in the EW universality class which suggests that $\nu_2$ in Eq.~(\ref{E:ContEqn}) is no longer zero in this dimension. They also observed mound formation in the WV simulations, instead of logarithmically flat surface, with $\alphaWV = 1$, $\betaWV = 1/4$, and $\zWV = 4$ after making use of the noise reduction technique. This implies that WV model is \emph{not} in the EW universality class in 2+1 dimensions. Moreover, the mounds tend to be of roughly equal size---an apparent deviation from being scale invariant.

Recently Haselwandter \emph{et al.}\cite{HaseVved07,HaseVved08} have shown that the unstable growth observed in WV model could be explained using renormalization group approach. They derived Eq.~(\ref{E:ContEqn}) from a master equation describing the increment of height at each lattice site according to the nearest-neighbor sites. By carefully choosing the regularization parameter upon taking a continuum limit, they were able to obtain the values of the coefficients $\nu_2$, $\nu_4$, $\lambda_{13}$ and $\lambda_{22}$. Under repeated RG transformations, these values flow differently in $d=1+1$ and $2+1$ dimensions. In particular the negativity of $\lambda_{13}$ leads to the change in the sign of the diffusion coefficient $\nu_2$ which eventually leads to the growth instability in the form of an array of islands of lateral size $\sim 2\pi\sqrt{2|\nu_4/\nu_2|}$. While their analysis gives a satisfactory account of the origin of mounds, their formulation is still appealed to an atypical regularization procedure with some dimensional dependency, and is not so conveniently extensible to analyze a more complicated lattice. In particular it is unclear whether the mechanism that gives rise to the growth instability is the property of the substrate dimension or of lattice geometry.

To investigate the surface morphologies of the select crystal lattices, we perform extensive simulations on chains with 100 to 250,000 elements. Due to differences in the number of atoms in a unit cell, these numbers translate to, e.g., the substrate of size $10\times 10$ to $500\times 500$ in the case of SC, and roughly $7\times 7$ to $353\times 353$ cells in the case of BCC(001). In all substrate sizes, the simulations are performed until they reach the time step beyond the point where the surface roughness $W$ saturates. This value ranges from $10^4$ UL for 100 elements, up to $10^7$ UL for 250,000 elements. Because of the cross-over behavior of the growth exponent $\beta$ at different timescales, its value changes slightly prior to the saturation time. 
%For each substrate size, we subdivide growth time into three regions: $1~\text{UL} \le t_1 \le t_\text{sat}/3$, $t_\text{sat}/3 \le t_2 \le 2t_\text{sat}/3$, and $2t_\text{sat}/3 \le t_3 \le t_\text{sat}$, where $t_\text{sat}$ denotes the saturation time for that particular substrate size. The value of $\beta$ is calculated by averaging $\beta$'s during $t_1$ and $t_2$.\footnote{The value of $\beta$ during timescale $t_3$ is disregarded because, according to Eq.~(\ref{E:scalingFn}), $\beta$ is defined during the early times.} 
The representative value of $\beta$ for a given crystal structure is obtained from the asymptotic value of the plot between $\beta(L)$ versus $1/L$ when $L\rightarrow \infty$. The roughness and dynamical exponents $\alpha$ and $z$ are computed from the slopes of $\text{log}( W_\text{sat})$ and $\text{log}(t_\text{sat})$ against the log of lateral substrate size $L$ respectively.

\begin{table}[htbp]
   \centering
   \begin{tabular}{@{} l @{\hspace{8pt}} c @{\hspace{8pt}} c @{\hspace{8pt}} c @{\hspace{8pt}} c @{}}
  		\toprule
    	  	\multicolumn{5}{c}{Das Sarma--Tamborenea model} \\
		\hline
		structure & $\alpha$ & $\beta$ & $z$ & mound \\
%		\hline
%		FCC(111) 	& $0.760\pm 0.036$	& $0.200\pm 0.030$	& $2.99\pm 0.21$ & yes(?)\\
%		SH		 	& $0.653\pm 0.010$	& $0.212\pm 0.042$	& $3.14\pm 0.09$ & no\\
%		iHCP			& $0.649\pm 0.002$	& $0.215\pm 0.044$	& $3.18\pm 0.09$ & no\\
%		FCC(001)		& $0.634\pm 0.015$	& $0.227\pm 0.049$	& $3.19\pm 0.17$ & no\\
%		SC			& $0.615\pm 0.018$	& $0.219\pm 0.035$	& $3.10\pm 0.21$ & no\\
%		BCC			& $0.570\pm 0.021$	& $0.237\pm 0.048$	& $3.12\pm 0.15$ & no\\
%		HCP			& $0.554\pm 0.026$	& $0.237\pm 0.042$	& $2.75\pm 0.20$ & no\\
		\hline
		FCC(111) 	& $0.76\pm 0.04$	& $0.20\pm 0.03$	& $3.3\pm 0.2$ & yes(?)\\
		SH		 	& $0.66\pm 0.01$	& $0.21\pm 0.05$	& $3.1\pm 0.1$ & no\\
		iHCP			& $0.65\pm 0.01$	& $0.22\pm 0.04$	& $3.2\pm 0.1$ & no\\
		FCC(001)		& $0.63\pm 0.02$	& $0.23\pm 0.05$	& $3.2\pm 0.1$ & no\\
		SC			& $0.62\pm 0.02$	& $0.22\pm 0.04$	& $3.1\pm 0.1$ & no\\
		BCC			& $0.57\pm 0.02$	& $0.24\pm 0.05$	& $3.1\pm 0.2$ & no\\
		HCP			& $0.52\pm 0.02$	& $0.24\pm 0.04$	& $2.8\pm 0.2$ & no\\
		\botrule
   \end{tabular}
   \caption{The critical exponents of the DT model for all lattice structures, sorted according to $\alpha$.}
   \label{tab:DTexponents}
\end{table}
\begin{table}[htbp]
   \centering
   \begin{tabular}{@{} l @{\hspace{8pt}} c @{\hspace{8pt}} c @{\hspace{8pt}} c @{\hspace{8pt}} c @{}}
   		\toprule
    	  	\multicolumn{4}{c}{Wolf--Villain model} \\
		\hline
		structure & $\alpha$ & $\beta$ & $z$ & mound \\
		\hline
%		SH		 	& $1.130\pm 0.015$	& $0.249\pm 0.040$	& $3.73\pm 0.30$ & yes\\
%		SC		 	& $0.935\pm 0.015$	& $0.233\pm 0.061$	& $3.60\pm 0.10$ & yes\\
%		iHCP			& $0.859\pm 0.016$	& $0.226\pm 0.024$	& $3.39\pm 0.27$ & yes\\
%		FCC(111)		& $0.833\pm 0.020$	& $0.199\pm 0.038$	& $3.55\pm 0.31$ & yes\\
%		FCC(001)		& $0.571\pm 0.032$	& $0.193\pm 0.009$	& $2.95\pm 0.24$ & no\\
%		BCC			& $0.455\pm 0.029$	& $0.198\pm 0.011$	& $2.11\pm 0.18$ & no\\
%		HCP			& $0.408\pm 0.031$	& $0.198\pm 0.014$	& $2.18\pm 0.18$ & no\\
		SH		 	& $1.12\pm 0.01$	& $0.25\pm 0.03$	& $3.9\pm 0.3$ & yes\\
		SC		 	& $0.94\pm 0.02$	& $0.23\pm 0.02$	& $3.6\pm 0.1$ & yes\\
		iHCP			& $0.86\pm 0.02$	& $0.23\pm 0.02$	& $3.5\pm 0.3$ & yes\\
		FCC(111)		& $0.83\pm 0.02$	& $0.20\pm 0.04$	& $3.6\pm 0.3$ & yes\\
		FCC(001)		& $0.57\pm 0.03$	& $0.19\pm 0.01$	& $3.0\pm 0.2$ & no\\
		BCC			& $0.42\pm 0.03$	& $0.20\pm 0.01$	& $1.9\pm 0.2$ & no\\
		HCP			& $0.37\pm 0.03$	& $0.20\pm 0.01$	& $2.2\pm 0.2$ & no\\
		\botrule
   \end{tabular}
   \caption{The critical exponents of the WV model for all lattice structures, sorted according to $\alpha$.}
   \label{tab:WVexponents}
\end{table}

Table~\ref{tab:DTexponents} and \ref{tab:WVexponents} show the values of the critical exponents for DT and WV models respectively. We find that the hyper-scaling relation $z = \alpha/\beta$ is \emph{not} generally respected within the simulation accuracy in both models when $\alpha$ is either too high or too low (such as HCP). Albeit some small variations, the values of the growth exponent $\beta$ agree with the predicted value ($\betaDTconv = 1/5$) from the continuum equation in the case of DT model across all lattice structures. The roughness exponent $\alpha$ and the dynamical exponent $z$ appear to be slightly less than those from the continuum predictions (except for $\alpha$ of FCC(111)). The exponents in the WV case, however, do not seem to conform to the theoretical prediction especially the dynamical exponent ($\zWVconv=2$) which ranges from approximately 2 to 4.
% REMOVE 02/18/10: Unlike in DT model, the growth exponent $\beta$ in WV model is substrate-size dependent which might indicate that the surface is logarithmically rough as dictated by the continuum equation.

The above discrepancy is removed by noticing the last column of Table~\ref{tab:WVexponents} which indicates the existence of mound-like morphologies on each substrate. In the case of BCC and HCP surfaces under WV diffusion rule, the surface front appears to be kinetically rough without any growth instability. We find a complete agreement between the values of the dynamical exponent from the simulations and that from the prediction of the continuum growth equation ($\zWVconv = 2$). (FCC(001) presents an exceptional case. We shall defer its discussion until Sec.~\ref{S:discussion}.)
Upon a closer examination of the last column of Table~\ref{tab:DTexponents} and \ref{tab:WVexponents}, we notice unstable mound-like morphologies for those structures with $\alpha > 0.66$ regardless of the diffusion model. (The reason for the question mark in the case of FCC(111) under DT model shall become evident at the end of Sec.~\ref{subsec:morph} and \ref{subsec:currents}.) The separation between mound and kinetically rough surfaces at a certain value of the roughness exponent has been previously observed in the experiment.\cite{LengPhanWillSarmBearJohn99} 
% Have to fix this a little bit. XXX
%The fact that the critical roughness exponent $\alpha_c$ having value greater than 0.66 is consistent with the absence of mound morphology from the simulations of any \emph{stable}, linear or nonlinear continuum growth equations of up to fourth order without ES barrier; the largest value of $\alpha$ in two spatial dimensions belongs to the MBE universality class with $\alpha = 2/3$.
The fact that the critical roughness exponent $\alpha_c$ having value greater than 0.66 is consistent with the presence of mound morphology from the simulations having a large enough value of $\alpha$ of the linear or nonlinear continuum growth equations and the MBE modellings of a SC lattice without ES barriers.\cite{SarmPunyToro00}

It should be noted that a ``rough'' surface indicates a large value of $W(L,t)$. Since we quantify ``roughness'' according to Eq.~(\ref{E:roughness}), ordered structures such as mounds or pyramids, tend to be ``rougher'' than scale-invariant, kinetically rough surfaces because the mound regions tend to be much higher, and the troughs of the hills much lower, than the average film height $\avg{h(t)}$. It is likely that a large value of the roughness exponent ($\alpha \approx 1$) would indicate mound morphology on a surface. When $\alpha = 1$, one obtains mounds with slope selection, \emph{i.e.}, mounds scale the same way as the lateral substrate dimension. For $\alpha > 1$ ($\alpha < 1$), mounds tend to grow (shrink) in size with a larger substrate. Surfaces that contain visible mounds, therefore, have a large value of $\alpha$, in contrast to kinetically rough surfaces of small $\alpha$ which appears flat upon taking the thermodynamic limit ($L\rightarrow \infty$).
% Since $\alpha$ measures how the saturated roughness scales with the substrate lateral size $W_\text{sat} \sim L^\alpha$, if one doubles the value of $L$, a surface with mounds would scale up ...
% XXX Find an explanation of why alpha is large for mound.

\section{Mounds and mechanism of mound formation}\label{S:currents}
It was well established that both step and kink Ehrlich--Schwoebel (ES) barriers could explain the formation of mound-like structures observed in many MBE growth experiments.\cite{ZuoWend97,YoonOhLee99,NeelMaroDouiErns03} The former prevents an atom on an upper terrace to hop down to a lower terrace, while the latter arises from a greater likelihood for an atom to move along the edge of a terrace toward a kink site than for it to move across a corner to a kink site.\cite{PierDorsEins99} Since then several authors\cite{BiehKinnKinzSchi99,ChatToroSarm01} have proposed a topologically induced probabilistic current known as step-edge diffusion (SED) current as an additional cause of mound formation. This type of current is physically similar to the energetically assisted kink ES current above. Their analysis, however, was based on a simple cubic structure. It is very unlikely that SED current is the only type of probabilistic, topological current in existence. An array of other geometrically more complicated crystalline structures could give rise to a new class of geometric, probabilistic current.

\subsection{Lattice-dependent mound morphology}\label{subsec:morph}
% XXX Figure of crystals with mounds prior to saturation.
In our simulations, we see mounds forming since an early stage of the growth naturally without any additional efforts for all structures with $\alpha > 0.66$. As time progresses, small mounds shift and coalesce into bigger ones, similar to what was reported from the WV results using the noise reduction technique.\cite{ChatToroSarm01,SarmChatToro02} The coarsening behavior with mounds of similar sizes during growth implies that the growth front is not scale invariant. The merging of mounds ends at $t_\text{sat}$ where the correlation length $\xi(t_\text{sat})$ is comparable to system's size, and only one mound (and one trough) remains. It is interesting to note that the growth and coarsening of mounds is not stationary; while a large mound subsumes smaller ones in order to grow, its tip does not stand still but shifts sideways in a series of disappearance and reemergence of a peak.

\begin{figure*}[htb]
	\centering
    \includegraphics[width=.49\textwidth]{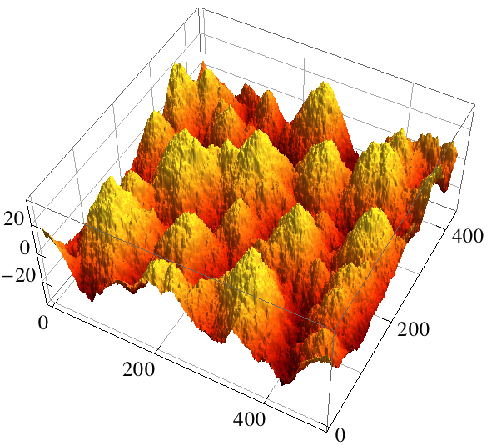}
    \includegraphics[width=.49\textwidth]{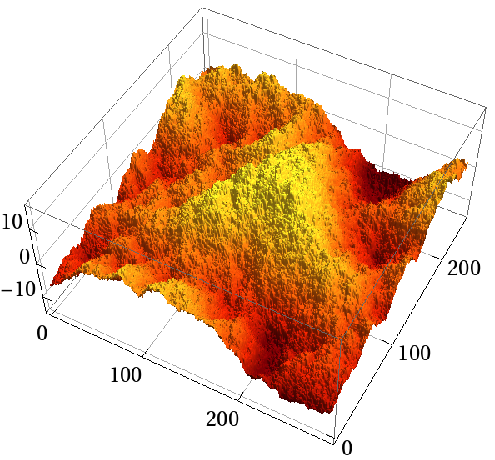}\\
    (a) \hspace{3.5in} (b)\\
    \vspace{.2in}
    \includegraphics[width=.48\textwidth]{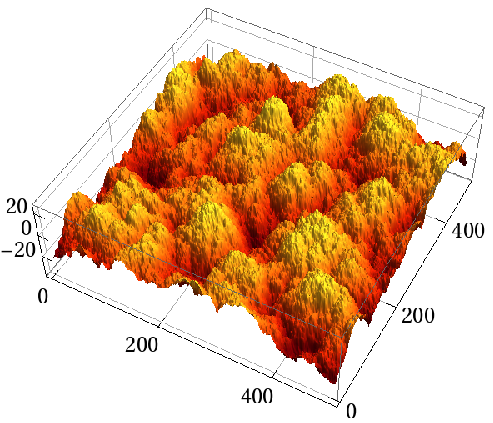}
    \includegraphics[width=.48\textwidth]{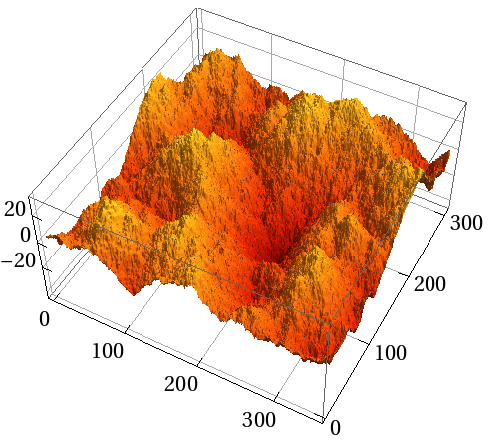}\\
    (c) \hspace{3.5in} (d)
    \caption{WV simulations of (a) SH, (b) FCC(111), (c) SC, and (d) iHCP lattices at $t=331132$ UL.}
    \label{fig:WVmounds}
\end{figure*}
Figure~\ref{fig:WVmounds} shows the surface morphologies prior to saturation times using WV model on chains with 250,000 elements. This model favors mounds because atoms tend to flow toward kink sites which are most likely to have the highest coordination numbers. DT model, on the other hand, is more inclined to generate a rough surface because adatoms generally stick to their original landings which already have high coordination numbers. Even within the same models, mounds do not assume the same form. Mounds found on SH and FCC(111) simulations exhibit strong faceted structures. In particular, FCC(111) simulations show a striking ensemble of triangular pyramids similar to many kinetic Monte Carlo (kMC) simulation results at low temperatures.\cite{OvesBogiLund99,AlbeMull05} Surfaces of SC and iHCP, on the other hand, only display semi-regular hillocks which do not reflect the underlining lattice structure. Three other lattice structures, namely FCC(001), BCC, and HCP, do not develop visible mounds within WV model.

% Figure of (a) FCC(111) DT, (b) SH
\begin{figure*}[htb]
	\centering
    \includegraphics[width=.49\textwidth]{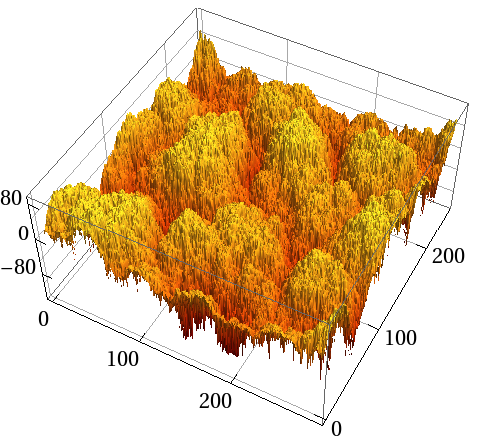}
    \includegraphics[width=.49\textwidth]{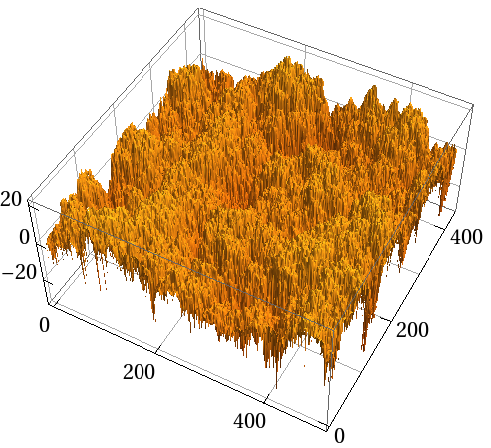}\\
    (a) \hspace{3.5in} (b)
    \caption{DT simulations of (a) FCC(111) and (b) SH at $t=331132$ UL.}
    \label{fig:FCCSHDT}
\end{figure*}
We do not see the development of mounds in most of our DT simulations on the same set of lattices---with a notable exception of FCC(111) as shown in Figure~\ref{fig:FCCSHDT}(a). While the surfaces of other crystalline structures appear to be statistically rough, the surface of FCC(111) show round, hemispherical mound morphologies, visually very similar to what was observed in DT model with ES barrier\cite{ChatToroSarm01} or in high-temperature kMC simulations\cite{SchiKohlReen00} on a simple cubic lattice. The second highest value of $\alpha$ among all lattices simulated using DT model is SH structure. Figure~\ref{fig:FCCSHDT}(b) shows the surface of SH prior to saturation without any apparent mound morphology. % XXX Check some fact here...
The surface of other lattices with a smaller value of $\alpha$ exhibits a statistically rough interface as traditionally expected from a standard DT simulation. On a closer examination, however, we find that the development of mounds on FCC(111) surface using DT model is quite different from the ones using WV model. Mounds, in this case, do not arise from island formations and coarsening of smaller mounds into larger ones. Initially the surface appears to be statistically rough. As time progresses, the regions which are later to become mounds, develop small cracks around them. The cracks then deepen, forming narrow troughs which gradually enlarge, splitting the original surface into many mounds. The perception of mound growth is in fact the deepening and broadening of the troughs. Finally at late times, small mounds start to merge by the progressive disappearance of troughs which separate them. We suspect that in this stage, the correlation length $\xi(t)$ dictates the size of each mound (which is comparable to the substrate size as the saturation time is reached).

\subsection{Topologically induced uphill currents}\label{subsec:currents}
To understand the mechanism of mound formation in both models, one should examine the area nearby a terrace edge which separates two flat regions. One commonly accepted explanation as to why an island nucleation leads to the formation of a large mound-like structure is due to the flow of atoms, on average, towards the mound region resulting in the net ``uphill'' current.\cite{Krug97,Golu97,MoldGolu00,PoliKrug00,Krug02} Without appealing to the use of ES barrier, we consider a topologically induced uphill current in the spirit of SED current. As anticipated, we find that all of the lattice structures that develop mounds appear to have SED current. To our surprise however, the conventional SED current is almost always cancelled by local downhill current. We also discover that SH and FCC possess yet another type of geometrically induced current. Unlike SED current which flows \emph{along} an edge of a terrace towards a kink site, the new current flows in the perpendicular direction towards the edge. We believe that the reason why this ``terrace diffusion'' (TD) current has never been observed is because in SC, where most simulations\cite{BiehKinnKinzSchi99,MurtCoop99} are based on, an equal and opposite current flows downhill. The uphill and downhill currents thus, on average, cancel each other leaving only SED current. It is worth mentioning that TD current is analogous to the edge ES current, whereas SED current is to kink ES current. The difference is that the edge ES current may occur on a one dimensional substrate, while TD current is only present on some crystalline lattices in two spatial dimensions.

%%% Figure SH100 showing SE current along [10-10]
\begin{figure}[htb]
	\centering
    \includegraphics[width=.5\textwidth]{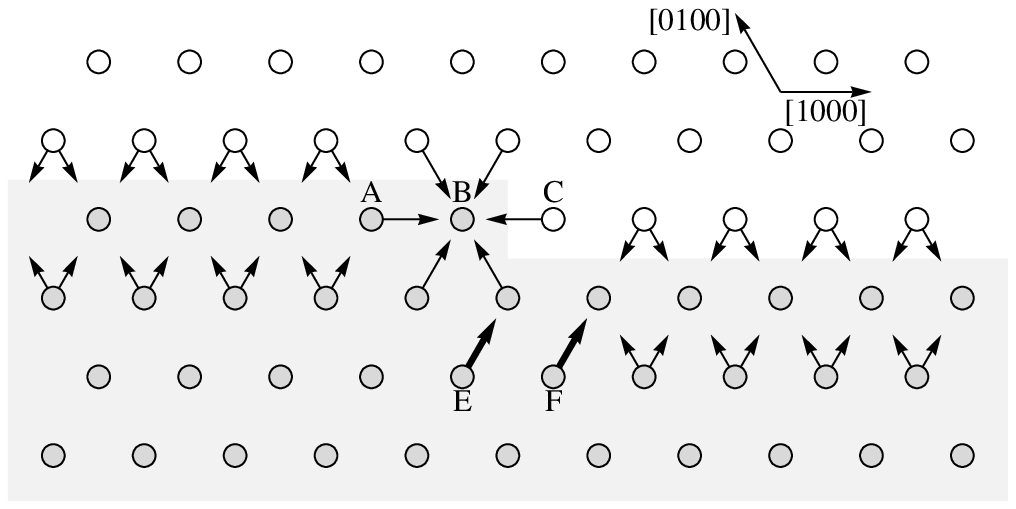}\\
    \caption{Local probabilistic currents near a step edge along [1000] direction of SH(0001) with a step dent. Shaded circles represent atoms on the lower terrace while light circles depict those on the upper terrace. An atom dropping on any lattice site will move, according to WV rule, to one of sites along the corresponding arrow. An atom will not move if it falls on a site without an arrow. An atom falling on site A, in particular, will be driven towards a kink site B producing a small SED current. This current is however cancelled by another downhill current from C to A. Global net currents are denoted by thick arrows.}
    \label{fig:SH100}
\end{figure}

To illustrate the difference between SED and TD, consider a step terrace lying along the [1000] direction of a SH(001) substrate as shown in Fig.~\ref{fig:SH100}. Atoms on the upper terrace are denoted by empty circles while those on the lower terrace are represented by shaded circles. According to WV model, a newly deposited atom which falls far from the edge of the terrace will not move. The one that falls within the proximity of the edge will advance along the direction(s) as shown by the arrow(s) in order to maximize its bondings. A site with two or more arrows indicates that there is an equal probability for an atom dropping on it to move in one of the allowed directions. Along the flat region away from the kink, atoms tend to move uphill as much as they move downhill resulting in a net zero flux. Near the kink site, we find that an uphill flux tends to occur more often. Note in particular that if an atom falls onto position A which situates on the edge, it will be attracted toward the kink position B creating a small SED current. Since WV diffusion rule only allows an atom to move to one of the nearest neighbors, the SED current only extends a distance of one atomic position. On average, however, a particle \emph{does not} tend to move uphill as a result of this current because there is another current flowing downhill in the opposite direction (from C to B) with the same strength. Nevertheless there is a net current in the uphill direction near the corner of the terrace edge at position E and F. It is not a SED current in the traditional sense since the direction of the flow is not along the edge but at an angle towards the corner. For the lack of a better word, we shall still refer to it as step-edge diffusion current because the current still appears in the neighborhood of a kink site and has a component parallel to a terrace edge.

%%% Figure SH120 showing TD current along [120]
\begin{figure}[htb]
	\centering
    \includegraphics[width=.45\textwidth]{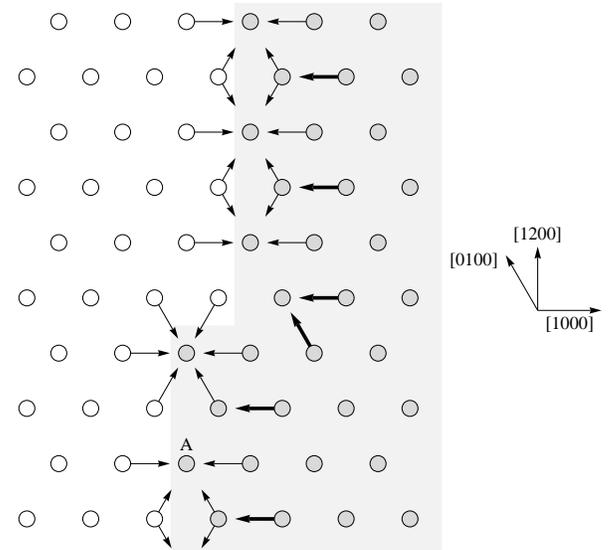}\\
    \caption{Local probabilistic currents near a step edge along [1200] direction of SH(0001). The upper terrace is on the left side while the lower terrace on the right. There is a net uphill terrace current acting along the [$\bar{1}$000] direction. Noncancelling currents are indicated by thick arrows.}
    \label{fig:SH120}
\end{figure}

The situation is even more perplexing for a step terrace along another high-symmetry direction, the [1200] direction, as shown in Fig.~\ref{fig:SH120}. Notice that an atom falling onto position A next to the kink site does not move toward the site. No (traditional) SED current exists along this direction. We, however, find a net noncancelling flux of currents (thick arrows) flowing perpendicular to the terrace edge in the uphill direction. This flux would serve to extend the base of the terrace in a future time step. (We shall discuss, in Sec.~\ref{S:discussion}, a notable exception of FCC(001) where a non-zero current does not lead to the formation of mounds.) To our knowledge, this type of topological current has never been reported in the literature. Near the corner, there also exists a SED current similar to those in Fig.~\ref{fig:SH100}. Table~\ref{tab:Currents} gives a summary of the type of currents along a given direction during the growth on SH, SC, iHCP, FCC(111), and FCC(001) surfaces. The upper terrace resides on the inside of the geometrical figures. It is interesting to note that FCC(111) simulations show very strong triangular pyramidal mounds oriented in the same direction, and never an inverted triangular version. We believe that this is due to the difference between the symmetry of the two types of currents; TD current is only three-fold symmetric while SED current has a six-fold symmetry. The preferred faces are oriented perpendicular to the directions of the TD currents, forming an upright triangular pyramid. The other structures whose surface has irregular mounds, namely SC and iHCP, are devoid of the TD currents. In addition to the edge ES current, TD current should cause an instability forming equilibrium faceting along some vicinal surfaces.

\begin{table}[htb]
	\centering
    \includegraphics[width=.49\textwidth]{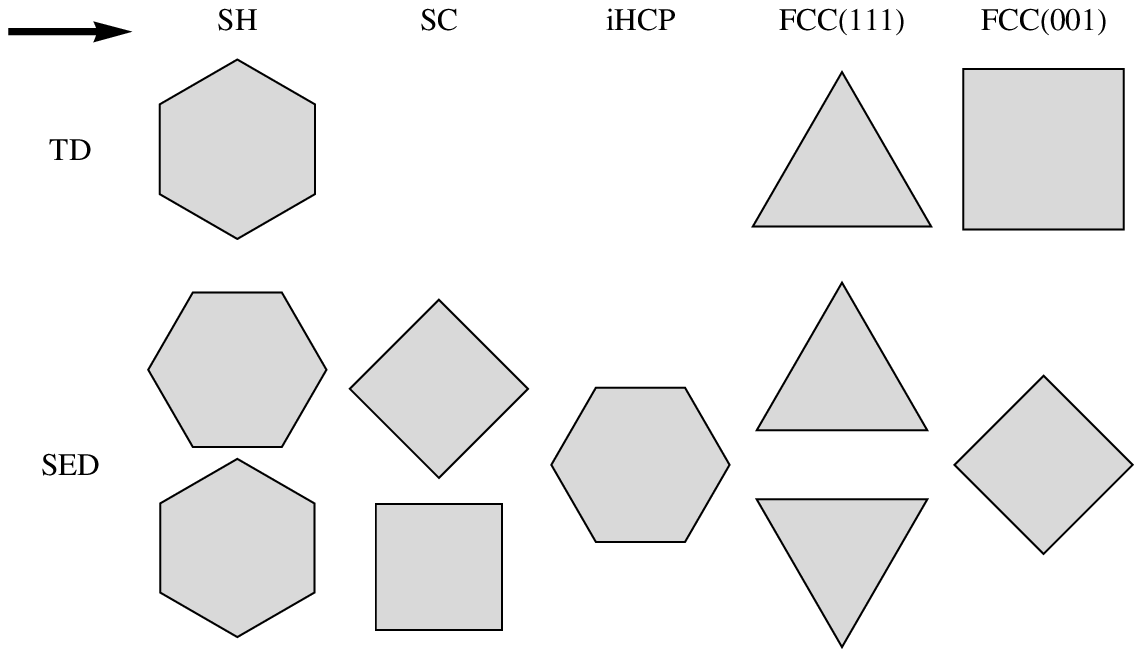}
    \caption{Edges of SH, SC, iHCP, FCC(111) and FCC(001) are shown where the corresponding terrace diffusion (TD) and step-edge diffusion (SED) currents are nonzero, calculated based on WV diffusion rule. Upper terraces are shown in gray. The horizontal direction, indicated by the arrow on the upper left hand corner, designates the [100] direction for SC and FCC(001), [1000] direction for SH and iHCP, and [1$\bar{1}$0] direction for FCC(111).}
    \label{tab:Currents}
\end{table}

%%% Figure of BCC
\begin{figure}[htb]
	\centering
    \includegraphics[width=.49\textwidth]{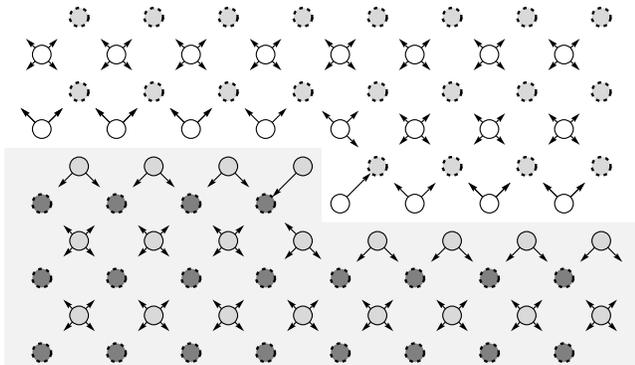}\\
    \caption{A step edge along the [100] direction on a BCC(001) plane separates the lower region (on the bottom of the figure) and the upper region (on the top). Three atomic layers are present with a lighter shade signifies a higher layer. Occupied lattice sites are represented by full circles while dashed circles denote unoccupied ones. Locally atoms tend to move away from the terrace edge.}
    \label{fig:BCC100}
\end{figure}

For BCC and HCP, mounds are not observed and neither type of current is present. Consider as an example the current consideration in the case of BCC(001), obeying WV model. Figure~\ref{fig:BCC100} shows three layers of atoms. Full circles signify occupied lattice sites. If an atom falls on one of these positions, it would have to move along a direction designated by one of the arrows towards an unoccupied site (dashed circle). Not only does a net current not exist, locally it flows perpendicularly \emph{away} from the terrace edge in both directions. Any atom deposits near the edge will likely be pushed away from it which implies that had an island been formed, its territory would not have been extended by this process. In addition, we also do not find any current flowing along the edge towards a kink site. In fact, if an atom falls exactly at the kink position, it will diffuse away from the kink. This counter-intuitive behavior arises from the fact that, for BCC, lattice sites along a terrace edge have the lowest coordination numbers. Moving towards these sites in the uphill direction from the bottom terrace would reduce the number of bonds, in contradiction with the WV diffusion rule. A closer inspection shows that atoms at the bottom of the edge already bond with those at the top. An atom which falls on either of these two rows adjacent to the edge can only roll away from the edge. The step edge in this case serves as a topological barrier preventing an atom to cross side. The same situation also happens in a HCP lattice whose surface is also mound-free.

%%% What happens to FCC(111) DT?
We end this section by giving a brief account of the DT simulation results. As discussed at the end of Sec.~\ref{subsec:morph}, no island formations are observed on the surface of these lattices, even in the case of FCC(111) where mounds are present. This is consistent with the fact that we do not find any non-zero uphill TD or SED currents on any structure in any direction. Other than FCC(111), all surfaces appear to be kinetically rough with early-time behavior following a power law.

\section{Discussion}\label{S:discussion}
% Why alpha = 0.66?
It is clear that the evolution of surface morphology depends not only on lattice dimension but also on material's crystal structure. In describing the growth of a lattice structure obeying a particular diffusion rule, the values (and signs) of the parameters ($\nu_2$, $\nu_4$, $\lambda_{13}$, $\lambda_{22}$ and $D$) in the associated continuum growth equation need to be adjusted accordingly. The growth morphology are primarily categorized into two classes: kinetically rough scale-invariant or unstable mounding surface. We find that the separation between these two growth regimes occurs at the roughness exponent $\alpha$ of around 0.66 regardless of the prescribed diffusion rule. Further analytical study is needed to explain the origin of this magic number. For kinetically rough surfaces, the dynamical scaling theory seems to give an accurate description of the behavior of the growth interface using power laws. On the contrary, for destabilized mounding morphologies, the growth needs to be described in terms of island nucleation and island coarsening. We shall leave the analysis of the dynamics of mound coarsening in limited mobility diffusion models for future work.\footnote{We refer the reader to Ref.~\onlinecite{Sieg98} for the analytical study of the effect of crystal anisotropy, slope selection and the dynamics of coarsening of mounds.}

% Why others don't see mounds in SC? -- small substrate, limited SE current
Contrary to Ref.~\onlinecite{ChatToroSarm01}, we are able to obtain mound morphology without any noise reduction technique. In our simulations we find that mounds are recognizable after its lateral size reaches about 100 atomic units. We do not see mounds comparable in size to theirs. Our supposition is that since we expect the parameters of the corresponding continuum growth equation to be substrate dependent, the uphill diffusion term $-|\nu_2| \nabla^2 h$ may overcome the Mullin-type diffusion term $-|\nu_4| \nabla^4 h$, which tends to suppress small fluctuations, at the length scale given by $l_c \sim \sqrt{|\nu_4/\nu_2|}$. This length scale $l_c$, in some crystal structure, may be larger than the attempted substrate simulation scale, thus, mounds may never be observed. In addition our mounds are much more irregular than the ones obtained using the noise reduced scheme. Our value of $\alpha$ for SC is very close to one which implies that mounds have a selective slope in agreement with Ref.~\onlinecite{SarmChatToro02}. We believe that the noise reduction technique, in most cases, serves to amplify the mound shape and is not a necessary scheme to produce mounds. 
% What happens to FCC(001)? Perhaps one needs NRT? What about FCC(111) DT?
An exception to this observation comes about in the case of FCC(001). Similar to SH case, we find both SED and TD currents acting along a terrace edge---albeit not both in the same direction---without any growth instability. We suspect, in this case, that either the strength of the uphill current is too weak in comparison to the randomness from the shot noise, or the substrate size is too limited to see the mound formation ($L < l_c$). More quantitative analysis of the current and a thorough investigation of the height-height correlation function are necessary to address this question.

% Propose terms which are responsible for TD current near the edge.
A few remarks are in order regarding the topological currents. Our observation leads us to believe that the mechanism of mound formation within our framework is due to both the kink SED current and the straight TD current. The latter serves as an extra role in enhancing the faceted structure of mounds on the surface where it exists. It is true that other, more complicated, edge shapes exist which could cause other geometrical currents. In a certain coarse-graining sense, dimples and pits can be generated by one kink/corner step similar to the ones in Fig.~\ref{fig:SH100}, \ref{fig:SH120} and \ref{fig:BCC100}. We still believe that these currents can be largely categorized into curvature-dependent versus straight terrace edge type of current. 
In his review article,\cite{Krug97} Krug argued that SED current induces $\vec{J}_\text{SED} \sim \vectsym{\nabla}\kappa(h)$, where $\kappa$ is the local curvature of $h(\vec{x},t)$ on the plane of the substrate. This results in $\sim \nabla^4 h$ in Eq.~(\ref{E:ContEqn}). Physically this term emerges as a result of the line tension due to the curvature of terrace edge. The TD current, on the other hand, appears even when the radius of curvature is infinite. Given $m \equiv |\hvect{n}\cdot\vectsym{\nabla} h|$ where $\hvect{n}$ defines the direction along which the current is active, TD current gives rise to the anti-diffusive flow along the uphill direction $\vec{J}_\text{TD} \sim \hvect{n}\, m/(1+m^2)$ which is approximately $\sim \hvect{n}\,m$ for a small surface slope.\cite{BaraStan95} Our findings suggest that both $\vec{J}_\text{SED}$ and $\vec{J}_\text{TD}$ only act along certain preferred directions according to the underlying lattice structure.% Other related articles
\footnote{An interested reader should consult, e.g., Ref.~\onlinecite{PierDorsEins99,MoldGolu00,PoliKrug00,Krug02} for detailed analyses of surface growth instability due to uphill currents and the relevant terms in the continuum growth equation.} 
(In the case of FCC(111), for example, the TD current may point along one of the following three directions: [112], [$\bar{2}$1$\bar{1}$] or [1$\bar{2}\bar{1}$].)  

Our simulations also suggest that, in most structures with surface growth instability, local current tends to flow towards the bottom of a terrace edge both from the upper and lower terraces. (See, e.g., Fig.~\ref{fig:SH100} and \ref{fig:SH120}.) On a kinetically rough surface such as that of BCC and HCP, we observe the local current which flows away from the bottom of a terrace edge, similar to what is seen in Fig.~\ref{fig:BCC100}. Although the effect of this current tends to average out on a larger scale, its existence gives rise to the current of the form $\vec{J}_\text{local} \sim \pm \hvect{n}(\hvect{n}\cdot\vectsym{\nabla})^2 h$ This translates to a new term proportional to $(\hvect{n}\cdot\vectsym{\nabla})^3 h$ in the continuum equation. (On a one dimensional substrate, this is simply $\partial^3 h/\partial x^3$.) This term has been previously neglected based on the rotation and inversion symmetry about the growth direction. Given an anisotropy of each lattice structure, we do not believe that the new term should be discarded from future investigations.

% Talk about FCC(111) DT case
Finally we see a different mound formation process on FCC(111) plane under DT diffusion rule. The traditional picture of island nucleation, followed by particle accretion and mound coarsening may not give an accurate description of the DT structural formation. Without any net uphill currents, we expect a completely different mechanism at work. Visually, mounds on FCC(111) plane do not possess up-down symmetry as the ones obtained using WV diffusion rule. This is, however, typical of DT growth morphology. We therefore still expect the DT terms ($-\nu_4\nabla^4 h$ and $\lambda_{22}\nabla^2(\vectsym{\nabla}h)^2$) to still be effective in the continuum growth equation. In light of the supposed anisotropy term that might be present, a complete understanding demands a more thorough theoretical investigation of the growth equation.

% Future plans
% Investigate mound growth scaling power.

\section{Conclusion}\label{S:conclusion}
Through large-scale Monte Carlo simulations, we have analyzed MBE growth of thin films on several lattice structures based on WV and DT models in 2+1 dimensions. We discovers that at the roughness exponent of around 0.66-0.76, the surface morphology of the film changes from being kinetically rough with power law scaling to quasi-regular mound-like structures. Without ES barrier, we attribute the morphological difference to the appearance of topologically induced, probabilistic particle currents. These currents not only arise from the line tension along the step edges separating several terraces of each mound in the form of SED current, they can also emerge perpendicular to flat straight terrace edges in the uphill direction in the form of TD current. The latter only manifests itself in SH and FCC lattices among several others that we have observed.

\begin{acknowledgments}
%The authors are grateful to Manit Klawtanong and Jindarat Yaemwong for their thoughtful input and valuable suggestions. We also would like to thank xxx for xxx. Funding from Thailand Center of Excellence in Physics under contract number xxx is acknowledged.
The authors would like to acknowledge fundings from Thailand Center of Excellence in Physics and CU Graduate School Thesis Grant. One of us (Kanjanaput) receives additional funding from the Development and Promotion of Science and Technology Talents Project.
\end{acknowledgments}

\bibliography{references}

\begin{thebibliography}{50}
\expandafter\ifx\csname natexlab\endcsname\relax\def\natexlab#1{#1}\fi
\expandafter\ifx\csname bibnamefont\endcsname\relax
  \def\bibnamefont#1{#1}\fi
\expandafter\ifx\csname bibfnamefont\endcsname\relax
  \def\bibfnamefont#1{#1}\fi
\expandafter\ifx\csname citenamefont\endcsname\relax
  \def\citenamefont#1{#1}\fi
\expandafter\ifx\csname url\endcsname\relax
  \def\url#1{\texttt{#1}}\fi
\expandafter\ifx\csname urlprefix\endcsname\relax\def\urlprefix{URL }\fi
\providecommand{\bibinfo}[2]{#2}
\providecommand{\eprint}[2][]{\url{#2}}

\bibitem[{\citenamefont{Chambers}(2000)}]{Cham00}
\bibinfo{author}{\bibfnamefont{S.~A.} \bibnamefont{Chambers}},
  \bibinfo{journal}{Surf. Sci. Repo.} \textbf{\bibinfo{volume}{39}},
  \bibinfo{pages}{105} (\bibinfo{year}{2000}).

\bibitem[{\citenamefont{Arthur}(2002)}]{Arth02}
\bibinfo{author}{\bibfnamefont{J.~R.} \bibnamefont{Arthur}},
  \bibinfo{journal}{Surf. Sci.} \textbf{\bibinfo{volume}{500}},
  \bibinfo{pages}{189} (\bibinfo{year}{2002}).

\bibitem[{\citenamefont{Barab\'asi and Stanley}(1995)}]{BaraStan95}
\bibinfo{author}{\bibfnamefont{A.-L.} \bibnamefont{Barab\'asi}}
  \bibnamefont{and} \bibinfo{author}{\bibfnamefont{H.~E.}
  \bibnamefont{Stanley}}, \emph{\bibinfo{title}{Fractal Concepts in Surface
  Growth}} (\bibinfo{publisher}{Cambridge University Press},
  \bibinfo{address}{Cambridge, MA}, \bibinfo{year}{1995}).

\bibitem[{\citenamefont{Pimpinelli and Villain}(1998)}]{PimpVill98}
\bibinfo{author}{\bibfnamefont{A.}~\bibnamefont{Pimpinelli}} \bibnamefont{and}
  \bibinfo{author}{\bibfnamefont{J.}~\bibnamefont{Villain}},
  \emph{\bibinfo{title}{Physics of Crystal Growth}}
  (\bibinfo{publisher}{Cambridge University Press},
  \bibinfo{address}{Cambridge, UK}, \bibinfo{year}{1998}).

\bibitem[{\citenamefont{Ernst et~al.}(1994)\citenamefont{Ernst, Fabre,
  Folkerts, and Lapujoulade}}]{ErnsFabrFolkLapu94}
\bibinfo{author}{\bibfnamefont{H.-J.} \bibnamefont{Ernst}},
  \bibinfo{author}{\bibfnamefont{F.}~\bibnamefont{Fabre}},
  \bibinfo{author}{\bibfnamefont{R.}~\bibnamefont{Folkerts}}, \bibnamefont{and}
  \bibinfo{author}{\bibfnamefont{J.}~\bibnamefont{Lapujoulade}},
  \bibinfo{journal}{Phys. Rev. Lett.} \textbf{\bibinfo{volume}{72}},
  \bibinfo{pages}{112} (\bibinfo{year}{1994}).

\bibitem[{\citenamefont{Coluci et~al.}(1998)\citenamefont{Coluci, Cotta,
  Mendon\c{c}a, I.-Landers, and {de Carvalho}}}]{ColuCottMendLandCarv98}
\bibinfo{author}{\bibfnamefont{V.~R.} \bibnamefont{Coluci}},
  \bibinfo{author}{\bibfnamefont{M.~A.} \bibnamefont{Cotta}},
  \bibinfo{author}{\bibfnamefont{C.~A.~C.} \bibnamefont{Mendon\c{c}a}},
  \bibinfo{author}{\bibfnamefont{K.~M.} \bibnamefont{I.-Landers}},
  \bibnamefont{and} \bibinfo{author}{\bibfnamefont{M.~M.~G.} \bibnamefont{{de
  Carvalho}}}, \bibinfo{journal}{Phys. Rev. B} \textbf{\bibinfo{volume}{58}},
  \bibinfo{pages}{1947} (\bibinfo{year}{1998}).

\bibitem[{\citenamefont{Nostrand et~al.}(1998)\citenamefont{Nostrand, Chey, and
  Cahill}}]{VanNostCheyCahi98}
\bibinfo{author}{\bibfnamefont{J.~E.~V.} \bibnamefont{Nostrand}},
  \bibinfo{author}{\bibfnamefont{S.~J.} \bibnamefont{Chey}}, \bibnamefont{and}
  \bibinfo{author}{\bibfnamefont{D.~G.} \bibnamefont{Cahill}},
  \bibinfo{journal}{Phys. Rev. B} \textbf{\bibinfo{volume}{57}},
  \bibinfo{pages}{12536} (\bibinfo{year}{1998}).

\bibitem[{\citenamefont{Kalff et~al.}(1999)\citenamefont{Kalff, {\v S}milauer,
  Comsa, and Michely}}]{KalfSmilComsMich99}
\bibinfo{author}{\bibfnamefont{M.}~\bibnamefont{Kalff}},
  \bibinfo{author}{\bibfnamefont{P.}~\bibnamefont{{\v S}milauer}},
  \bibinfo{author}{\bibfnamefont{G.}~\bibnamefont{Comsa}}, \bibnamefont{and}
  \bibinfo{author}{\bibfnamefont{T.}~\bibnamefont{Michely}},
  \bibinfo{journal}{Surf. Sci.} \textbf{\bibinfo{volume}{426}},
  \bibinfo{pages}{L447} (\bibinfo{year}{1999}).

\bibitem[{\citenamefont{Finnie and Homma}(1999)}]{FinnHomm99}
\bibinfo{author}{\bibfnamefont{P.}~\bibnamefont{Finnie}} \bibnamefont{and}
  \bibinfo{author}{\bibfnamefont{Y.}~\bibnamefont{Homma}},
  \bibinfo{journal}{Phys. Rev. B} \textbf{\bibinfo{volume}{59}},
  \bibinfo{pages}{15240} (\bibinfo{year}{1999}).

\bibitem[{\citenamefont{Fauster et~al.}(2000)\citenamefont{Fauster, Reu\ss,
  Shumay, and Weinelt}}]{FausReusShumWeinTheiGold00}
\bibinfo{author}{\bibfnamefont{T.}~\bibnamefont{Fauster}},
  \bibinfo{author}{\bibfnamefont{C.}~\bibnamefont{Reu\ss}},
  \bibinfo{author}{\bibfnamefont{I.~L.} \bibnamefont{Shumay}},
  \bibnamefont{and} \bibinfo{author}{\bibfnamefont{M.}~\bibnamefont{Weinelt}},
  \bibinfo{journal}{Phys. Rev. B} \textbf{\bibinfo{volume}{61}},
  \bibinfo{pages}{16168} (\bibinfo{year}{2000}).

\bibitem[{\citenamefont{N\'eel et~al.}(2003)\citenamefont{N\'eel, Maroutian,
  Douillard, and Ernst}}]{NeelMaroDouiErns03}
\bibinfo{author}{\bibfnamefont{N.}~\bibnamefont{N\'eel}},
  \bibinfo{author}{\bibfnamefont{T.}~\bibnamefont{Maroutian}},
  \bibinfo{author}{\bibfnamefont{L.}~\bibnamefont{Douillard}},
  \bibnamefont{and} \bibinfo{author}{\bibfnamefont{H.-J.} \bibnamefont{Ernst}},
  \bibinfo{journal}{Phys. Rev. Lett.} \textbf{\bibinfo{volume}{91}},
  \bibinfo{pages}{226103} (\bibinfo{year}{2003}).

\bibitem[{\citenamefont{Ehrlich and Hudda}(1966)}]{EhrlHudd66}
\bibinfo{author}{\bibfnamefont{G.}~\bibnamefont{Ehrlich}} \bibnamefont{and}
  \bibinfo{author}{\bibfnamefont{F.~G.} \bibnamefont{Hudda}},
  \bibinfo{journal}{J. Chem. Phys.} \textbf{\bibinfo{volume}{44}},
  \bibinfo{pages}{1039} (\bibinfo{year}{1966}).

\bibitem[{\citenamefont{Schwoebel and Shipsey}(1966)}]{SchwShip66}
\bibinfo{author}{\bibfnamefont{R.~L.} \bibnamefont{Schwoebel}}
  \bibnamefont{and} \bibinfo{author}{\bibfnamefont{E.~J.}
  \bibnamefont{Shipsey}}, \bibinfo{journal}{J. Appl. Phys.}
  \textbf{\bibinfo{volume}{37}}, \bibinfo{pages}{3682} (\bibinfo{year}{1966}).

\bibitem[{\citenamefont{Schwoebel}(1969)}]{SchwShip69}
\bibinfo{author}{\bibfnamefont{R.~L.} \bibnamefont{Schwoebel}},
  \bibinfo{journal}{J. Appl. Phys.} \textbf{\bibinfo{volume}{40}},
  \bibinfo{pages}{614} (\bibinfo{year}{1969}).

\bibitem[{\citenamefont{Villain}(1991)}]{Vill91}
\bibinfo{author}{\bibfnamefont{J.}~\bibnamefont{Villain}}, \bibinfo{journal}{J.
  Phys. I} \textbf{\bibinfo{volume}{1}}, \bibinfo{pages}{19}
  (\bibinfo{year}{1991}).

\bibitem[{\citenamefont{Burton et~al.}(1951)\citenamefont{Burton, Cabrera, and
  Frank}}]{BurtCabrFran51}
\bibinfo{author}{\bibfnamefont{W.~K.} \bibnamefont{Burton}},
  \bibinfo{author}{\bibfnamefont{N.}~\bibnamefont{Cabrera}}, \bibnamefont{and}
  \bibinfo{author}{\bibfnamefont{F.~C.} \bibnamefont{Frank}},
  \bibinfo{journal}{Phil. Trans. R. Soc. (Lond.) A}
  \textbf{\bibinfo{volume}{243}}, \bibinfo{pages}{299} (\bibinfo{year}{1951}).

\bibitem[{\citenamefont{Krug}(1997)}]{Krug97}
\bibinfo{author}{\bibfnamefont{J.}~\bibnamefont{Krug}},
  \bibinfo{journal}{Advances in Physics} \textbf{\bibinfo{volume}{46}},
  \bibinfo{pages}{139} (\bibinfo{year}{1997}).

\bibitem[{\citenamefont{Politi and Villain}(1996)}]{PoliVill96}
\bibinfo{author}{\bibfnamefont{P.}~\bibnamefont{Politi}} \bibnamefont{and}
  \bibinfo{author}{\bibfnamefont{J.}~\bibnamefont{Villain}},
  \bibinfo{journal}{Phys. Rev. B} \textbf{\bibinfo{volume}{54}},
  \bibinfo{pages}{5114} (\bibinfo{year}{1996}).

\bibitem[{\citenamefont{Camarero et~al.}(1999)\citenamefont{Camarero, Cros,
  Capit\'an, \'Alvarez, Ferrer, Ni{\~n}o, Prieto, Ferr\'on, de~Parga, Gallego
  et~al.}}]{CamaCrosCapiAlvaFerr99}
\bibinfo{author}{\bibfnamefont{J.}~\bibnamefont{Camarero}},
  \bibinfo{author}{\bibfnamefont{V.}~\bibnamefont{Cros}},
  \bibinfo{author}{\bibfnamefont{M.~J.} \bibnamefont{Capit\'an}},
  \bibinfo{author}{\bibfnamefont{J.}~\bibnamefont{\'Alvarez}},
  \bibinfo{author}{\bibfnamefont{S.}~\bibnamefont{Ferrer}},
  \bibinfo{author}{\bibfnamefont{M.~A.} \bibnamefont{Ni{\~n}o}},
  \bibinfo{author}{\bibfnamefont{J.~E.} \bibnamefont{Prieto}},
  \bibinfo{author}{\bibfnamefont{J.}~\bibnamefont{Ferr\'on}},
  \bibinfo{author}{\bibfnamefont{A.~L.~V.} \bibnamefont{de~Parga}},
  \bibinfo{author}{\bibfnamefont{J.~M.} \bibnamefont{Gallego}},
  \bibnamefont{et~al.}, \bibinfo{journal}{Appl. Phys. A}
  \textbf{\bibinfo{volume}{69}}, \bibinfo{pages}{553} (\bibinfo{year}{1999}).

\bibitem[{\citenamefont{Schinzer et~al.}(2000)\citenamefont{Schinzer, K\"ohler,
  and Reents}}]{SchiKohlReen00}
\bibinfo{author}{\bibfnamefont{S.}~\bibnamefont{Schinzer}},
  \bibinfo{author}{\bibfnamefont{S.}~\bibnamefont{K\"ohler}}, \bibnamefont{and}
  \bibinfo{author}{\bibfnamefont{G.}~\bibnamefont{Reents}},
  \bibinfo{journal}{Eur. Phys. J. B} \textbf{\bibinfo{volume}{15}}
  (\bibinfo{year}{2000}).

\bibitem[{\citenamefont{Liu and Shen}(2005)}]{LiuShen05}
\bibinfo{author}{\bibfnamefont{Z.-J.} \bibnamefont{Liu}} \bibnamefont{and}
  \bibinfo{author}{\bibfnamefont{Y.~G.} \bibnamefont{Shen}},
  \bibinfo{journal}{Surf. Sci.} \textbf{\bibinfo{volume}{595}},
  \bibinfo{pages}{20} (\bibinfo{year}{2005}).

\bibitem[{\citenamefont{Murty and Cooper}(1999)}]{MurtCoop99}
\bibinfo{author}{\bibfnamefont{M.~V.~R.} \bibnamefont{Murty}} \bibnamefont{and}
  \bibinfo{author}{\bibfnamefont{B.~H.} \bibnamefont{Cooper}},
  \bibinfo{journal}{Phys. Rev. Lett.} \textbf{\bibinfo{volume}{83}},
  \bibinfo{pages}{352} (\bibinfo{year}{1999}).

\bibitem[{\citenamefont{Pierre-Louis et~al.}(1999)\citenamefont{Pierre-Louis,
  D'Orsogna, and Einstein}}]{PierDorsEins99}
\bibinfo{author}{\bibfnamefont{O.}~\bibnamefont{Pierre-Louis}},
  \bibinfo{author}{\bibfnamefont{M.~R.} \bibnamefont{D'Orsogna}},
  \bibnamefont{and} \bibinfo{author}{\bibfnamefont{T.~L.}
  \bibnamefont{Einstein}}, \bibinfo{journal}{Phys. Rev. Lett.}
  \textbf{\bibinfo{volume}{82}}, \bibinfo{pages}{3661} (\bibinfo{year}{1999}).

\bibitem[{\citenamefont{Chatraphorn et~al.}(2001)\citenamefont{Chatraphorn,
  Toroczkai, and {Das Sarma}}}]{ChatToroSarm01}
\bibinfo{author}{\bibfnamefont{P.~P.} \bibnamefont{Chatraphorn}},
  \bibinfo{author}{\bibfnamefont{Z.}~\bibnamefont{Toroczkai}},
  \bibnamefont{and} \bibinfo{author}{\bibfnamefont{S.}~\bibnamefont{{Das
  Sarma}}}, \bibinfo{journal}{Phys. Rev. B} \textbf{\bibinfo{volume}{64}},
  \bibinfo{pages}{205407} (\bibinfo{year}{2001}).

\bibitem[{\citenamefont{Rangdee and Chatraphorn}(2006)}]{RangChat06}
\bibinfo{author}{\bibfnamefont{R.}~\bibnamefont{Rangdee}} \bibnamefont{and}
  \bibinfo{author}{\bibfnamefont{P.}~\bibnamefont{Chatraphorn}},
  \bibinfo{journal}{Surf. Sci.} \textbf{\bibinfo{volume}{600}},
  \bibinfo{pages}{914} (\bibinfo{year}{2006}).

\bibitem[{\citenamefont{{Das Sarma} et~al.}(2000)\citenamefont{{Das Sarma},
  Punyindu, and Toroczkai}}]{SarmPunyToro00}
\bibinfo{author}{\bibfnamefont{S.}~\bibnamefont{{Das Sarma}}},
  \bibinfo{author}{\bibfnamefont{P.}~\bibnamefont{Punyindu}}, \bibnamefont{and}
  \bibinfo{author}{\bibfnamefont{Z.}~\bibnamefont{Toroczkai}},
  \bibinfo{journal}{Surf. Sci.} \textbf{\bibinfo{volume}{457}},
  \bibinfo{pages}{L369} (\bibinfo{year}{2000}).

\bibitem[{\citenamefont{Chatraphorn and {Das Sarma}}(2002)}]{ChatSarm02}
\bibinfo{author}{\bibfnamefont{P.~P.} \bibnamefont{Chatraphorn}}
  \bibnamefont{and} \bibinfo{author}{\bibfnamefont{S.}~\bibnamefont{{Das
  Sarma}}}, \bibinfo{journal}{Phys. Rev. E} \textbf{\bibinfo{volume}{66}},
  \bibinfo{pages}{041601} (\bibinfo{year}{2002}).

\bibitem[{\citenamefont{{Das Sarma} et~al.}(2002)\citenamefont{{Das Sarma},
  Chatraphorn, and Toroczkai}}]{SarmChatToro02}
\bibinfo{author}{\bibfnamefont{S.}~\bibnamefont{{Das Sarma}}},
  \bibinfo{author}{\bibfnamefont{P.~P.} \bibnamefont{Chatraphorn}},
  \bibnamefont{and}
  \bibinfo{author}{\bibfnamefont{Z.}~\bibnamefont{Toroczkai}},
  \bibinfo{journal}{Phys. Rev. E} \textbf{\bibinfo{volume}{65}},
  \bibinfo{pages}{036144} (\bibinfo{year}{2002}).

\bibitem[{\citenamefont{Oliveira and {Aar{\~a}o Reis}}(2007)}]{OlivReis07}
\bibinfo{author}{\bibfnamefont{T.~J.} \bibnamefont{Oliveira}} \bibnamefont{and}
  \bibinfo{author}{\bibfnamefont{F.~D.~A.} \bibnamefont{{Aar{\~a}o Reis}}},
  \bibinfo{journal}{Phys. Rev. E} \textbf{\bibinfo{volume}{76}},
  \bibinfo{pages}{061601} (\bibinfo{year}{2007}).

\bibitem[{\citenamefont{Wolf and Villain}(1990)}]{WolfVill90}
\bibinfo{author}{\bibfnamefont{D.~E.} \bibnamefont{Wolf}} \bibnamefont{and}
  \bibinfo{author}{\bibfnamefont{J.}~\bibnamefont{Villain}},
  \bibinfo{journal}{Europhys. Lett.} \textbf{\bibinfo{volume}{13}},
  \bibinfo{pages}{389} (\bibinfo{year}{1990}).

\bibitem[{\citenamefont{{Das Sarma} and Tamborenea}(1991)}]{SarmTamb91}
\bibinfo{author}{\bibfnamefont{S.}~\bibnamefont{{Das Sarma}}} \bibnamefont{and}
  \bibinfo{author}{\bibfnamefont{P.~I.} \bibnamefont{Tamborenea}},
  \bibinfo{journal}{Phys. Rev. Lett.} \textbf{\bibinfo{volume}{66}},
  \bibinfo{pages}{325} (\bibinfo{year}{1991}).

\bibitem[{\citenamefont{Edwards and Wilkinson}(1982)}]{EdwaWilk82}
\bibinfo{author}{\bibfnamefont{S.~F.} \bibnamefont{Edwards}} \bibnamefont{and}
  \bibinfo{author}{\bibfnamefont{D.~R.} \bibnamefont{Wilkinson}},
  \bibinfo{journal}{Proc. Royal Soc. London A} \textbf{\bibinfo{volume}{381}},
  \bibinfo{pages}{17} (\bibinfo{year}{1982}).

\bibitem[{\citenamefont{Newman and Barkema}(1999)}]{NewmBark01}
\bibinfo{author}{\bibfnamefont{M.~E.~J.} \bibnamefont{Newman}}
  \bibnamefont{and} \bibinfo{author}{\bibfnamefont{G.~T.}
  \bibnamefont{Barkema}}, \emph{\bibinfo{title}{Monte Carlo Methods in
  Statistical Physics}} (\bibinfo{publisher}{Clarendon Press},
  \bibinfo{address}{Oxford, UK}, \bibinfo{year}{1999}).

\bibitem[{\citenamefont{Kardar et~al.}(1986)\citenamefont{Kardar, Parisi, and
  Zhang}}]{KardPariZhan86}
\bibinfo{author}{\bibfnamefont{M.}~\bibnamefont{Kardar}},
  \bibinfo{author}{\bibfnamefont{G.}~\bibnamefont{Parisi}}, \bibnamefont{and}
  \bibinfo{author}{\bibfnamefont{Y.-C.} \bibnamefont{Zhang}},
  \bibinfo{journal}{Phys. Rev. Lett.} \textbf{\bibinfo{volume}{56}},
  \bibinfo{pages}{889} (\bibinfo{year}{1986}).

\bibitem[{\citenamefont{Eden}(1961)}]{Eden61}
\bibinfo{author}{\bibfnamefont{M.}~\bibnamefont{Eden}}, in
  \emph{\bibinfo{booktitle}{Proceedings of the Fourth Berkeley Symposium on
  Mathematical Statistics and Probability}}, edited by
  \bibinfo{editor}{\bibfnamefont{F.}~\bibnamefont{Neyman}},
  \bibinfo{organization}{The University of California at Berkeley}
  (\bibinfo{publisher}{Univ. of Calif. Press}, \bibinfo{year}{1961}),
  vol.~\bibinfo{volume}{IV}, pp. \bibinfo{pages}{223--39}.

\bibitem[{\citenamefont{Witten and Sander}(1981)}]{WittSand81}
\bibinfo{author}{\bibfnamefont{T.~A.} \bibnamefont{Witten}} \bibnamefont{and}
  \bibinfo{author}{\bibfnamefont{L.~M.} \bibnamefont{Sander}},
  \bibinfo{journal}{Phys. Rev. Lett.} \textbf{\bibinfo{volume}{47}},
  \bibinfo{pages}{1400} (\bibinfo{year}{1981}).

\bibitem[{\citenamefont{Bazant et~al.}(2003)\citenamefont{Bazant, Choi, and
  Davidovitch}}]{BazaChoiDavi03}
\bibinfo{author}{\bibfnamefont{M.~Z.} \bibnamefont{Bazant}},
  \bibinfo{author}{\bibfnamefont{J.}~\bibnamefont{Choi}}, \bibnamefont{and}
  \bibinfo{author}{\bibfnamefont{B.}~\bibnamefont{Davidovitch}},
  \bibinfo{journal}{Phys. Rev. Lett.} \textbf{\bibinfo{volume}{91}},
  \bibinfo{pages}{045503} (\bibinfo{year}{2003}).

\bibitem[{\citenamefont{Haselwandter and Vvedensky}(2007)}]{HaseVved07}
\bibinfo{author}{\bibfnamefont{C.~A.} \bibnamefont{Haselwandter}}
  \bibnamefont{and} \bibinfo{author}{\bibfnamefont{D.~D.}
  \bibnamefont{Vvedensky}}, \bibinfo{journal}{Phys. Rev. Lett.}
  \textbf{\bibinfo{volume}{98}}, \bibinfo{pages}{046102}
  (\bibinfo{year}{2007}).

\bibitem[{\citenamefont{Haselwandter and Vvedensky}(2008)}]{HaseVved08}
\bibinfo{author}{\bibfnamefont{C.~A.} \bibnamefont{Haselwandter}}
  \bibnamefont{and} \bibinfo{author}{\bibfnamefont{D.~D.}
  \bibnamefont{Vvedensky}}, \bibinfo{journal}{Phys. Rev. E}
  \textbf{\bibinfo{volume}{77}}, \bibinfo{pages}{061129}
  (\bibinfo{year}{2008}).

\bibitem[{\citenamefont{Lengel et~al.}(1999)\citenamefont{Lengel, Phaneuf,
  Williams, {Das Sarma}, Beard, and Johnson}}]{LengPhanWillSarmBearJohn99}
\bibinfo{author}{\bibfnamefont{G.}~\bibnamefont{Lengel}},
  \bibinfo{author}{\bibfnamefont{R.~J.} \bibnamefont{Phaneuf}},
  \bibinfo{author}{\bibfnamefont{E.~D.} \bibnamefont{Williams}},
  \bibinfo{author}{\bibfnamefont{S.}~\bibnamefont{{Das Sarma}}},
  \bibinfo{author}{\bibfnamefont{W.}~\bibnamefont{Beard}}, \bibnamefont{and}
  \bibinfo{author}{\bibfnamefont{F.~G.} \bibnamefont{Johnson}},
  \bibinfo{journal}{Phys. Rev. B} \textbf{\bibinfo{volume}{60}},
  \bibinfo{pages}{R8469} (\bibinfo{year}{1999}).

\bibitem[{\citenamefont{Zuo and Wendelken}(1997)}]{ZuoWend97}
\bibinfo{author}{\bibfnamefont{J.-K.} \bibnamefont{Zuo}} \bibnamefont{and}
  \bibinfo{author}{\bibfnamefont{J.~F.} \bibnamefont{Wendelken}},
  \bibinfo{journal}{Phys. Rev. Lett.} \textbf{\bibinfo{volume}{78}},
  \bibinfo{pages}{2791} (\bibinfo{year}{1997}).

\bibitem[{\citenamefont{Yoon et~al.}(1999)\citenamefont{Yoon, Oh, and
  Lee}}]{YoonOhLee99}
\bibinfo{author}{\bibfnamefont{J.-G.} \bibnamefont{Yoon}},
  \bibinfo{author}{\bibfnamefont{H.~K.} \bibnamefont{Oh}}, \bibnamefont{and}
  \bibinfo{author}{\bibfnamefont{S.~J.} \bibnamefont{Lee}},
  \bibinfo{journal}{Phys. Rev. B} \textbf{\bibinfo{volume}{60}},
  \bibinfo{pages}{2839} (\bibinfo{year}{1999}).

\bibitem[{\citenamefont{Biehl et~al.}(1999)\citenamefont{Biehl, Kinne, Kinzel,
  and Schinzer}}]{BiehKinnKinzSchi99}
\bibinfo{author}{\bibfnamefont{M.}~\bibnamefont{Biehl}},
  \bibinfo{author}{\bibfnamefont{M.}~\bibnamefont{Kinne}},
  \bibinfo{author}{\bibfnamefont{W.}~\bibnamefont{Kinzel}}, \bibnamefont{and}
  \bibinfo{author}{\bibfnamefont{S.}~\bibnamefont{Schinzer}},
  \bibinfo{journal}{Comput. Phys. Comm.} \textbf{\bibinfo{volume}{121--122}},
  \bibinfo{pages}{347} (\bibinfo{year}{1999}).

\bibitem[{\citenamefont{Ovesson et~al.}(1999)\citenamefont{Ovesson, Bogicevic,
  and Lundqvist}}]{OvesBogiLund99}
\bibinfo{author}{\bibfnamefont{S.}~\bibnamefont{Ovesson}},
  \bibinfo{author}{\bibfnamefont{A.}~\bibnamefont{Bogicevic}},
  \bibnamefont{and} \bibinfo{author}{\bibfnamefont{B.~I.}
  \bibnamefont{Lundqvist}}, \bibinfo{journal}{Phys. Rev. Lett.}
  \textbf{\bibinfo{volume}{83}}, \bibinfo{pages}{2608} (\bibinfo{year}{1999}).

\bibitem[{\citenamefont{Albe and M\"uller}(2005)}]{AlbeMull05}
\bibinfo{author}{\bibfnamefont{K.}~\bibnamefont{Albe}} \bibnamefont{and}
  \bibinfo{author}{\bibfnamefont{M.}~\bibnamefont{M\"uller}},
  \bibinfo{journal}{Int. Numer. Math.} \textbf{\bibinfo{volume}{149}},
  \bibinfo{pages}{19} (\bibinfo{year}{2005}).

\bibitem[{\citenamefont{Golubovi{\'{c}}}(1997)}]{Golu97}
\bibinfo{author}{\bibfnamefont{L.}~\bibnamefont{Golubovi{\'{c}}}},
  \bibinfo{journal}{Phys. Rev. Lett.} \textbf{\bibinfo{volume}{78}},
  \bibinfo{pages}{90} (\bibinfo{year}{1997}).

\bibitem[{\citenamefont{Moldovan and Golubovi{\'{c}}}(2000)}]{MoldGolu00}
\bibinfo{author}{\bibfnamefont{D.}~\bibnamefont{Moldovan}} \bibnamefont{and}
  \bibinfo{author}{\bibfnamefont{L.}~\bibnamefont{Golubovi{\'{c}}}},
  \bibinfo{journal}{Phys. Rev. E} \textbf{\bibinfo{volume}{61}},
  \bibinfo{pages}{6190} (\bibinfo{year}{2000}).

\bibitem[{\citenamefont{Politi and Krug}(2000)}]{PoliKrug00}
\bibinfo{author}{\bibfnamefont{P.}~\bibnamefont{Politi}} \bibnamefont{and}
  \bibinfo{author}{\bibfnamefont{J.}~\bibnamefont{Krug}},
  \bibinfo{journal}{Surf. Sci.} \textbf{\bibinfo{volume}{446}},
  \bibinfo{pages}{89} (\bibinfo{year}{2000}).

\bibitem[{\citenamefont{Krug}(2002)}]{Krug02}
\bibinfo{author}{\bibfnamefont{J.}~\bibnamefont{Krug}},
  \bibinfo{journal}{Physica A} \textbf{\bibinfo{volume}{313}},
  \bibinfo{pages}{47} (\bibinfo{year}{2002}).

\bibitem[{\citenamefont{Siegert}(1998)}]{Sieg98}
\bibinfo{author}{\bibfnamefont{M.}~\bibnamefont{Siegert}},
  \bibinfo{journal}{Phys. Rev. Lett.} \textbf{\bibinfo{volume}{81}},
  \bibinfo{pages}{5481} (\bibinfo{year}{1998}).

\end{thebibliography}

\end{document}